# Understanding infection risks of COVID-19 in the city: an investigation of infected neighborhoods in Wuhan

Weipan Xu, Ying Li, Xun Li

## Abstract

During the COVID-19 pandemic, built environments in dense urban settings become major sources of infection. This study tests the difference of demographics and surrounding built environments across high-, medium- and low-infection neighborhoods, to inform the high-risk areas in the city. We found that high-infection neighborhoods own a higher ratio of aged population than other neighborhoods on average. However, it shows no statistical difference in terms of population density. Additionally, high-infection neighborhoods are closer to high-risk built environments than the others. In a walking distance, they also can access more of the high-risk built environments except for the wholesale markets and shopping malls. These findings advise policy-makers to deploy social distancing measures in precision, regulating the access of high-risk facilities to mitigate the impacts of COVID-19.

## Introduction

Social distancing is widely adopted to slow down the spread of COVID-19[1,2]. The epi-center, Wuhan, has successfully contained the infectious disease through effective measures, significantly reducing the daily confirmed cases from 1500-2000 at its peak to 10 cases or less a day[3]. Cities that managed to contain COVID-19 would be keen to lift the measures as soon as possible, reviving the economic activities and social interactions back to the pre-quarantine level. Therefore, major cities of China have gradually eased the social distancing measures since late February[4]. In a dense urban setting, if certain effective measures were lifted in areas that tend to spread the virus, there could be risks of a second major outbreak, compromising the hard-earned success. In this regard, understanding which types of built environments (BEs) in the city that could be high-risk areas of infection is of significant importance for the policy-makers to implement social distancing measures in precision.

To understand the high-risk BE in the city, we should first identify the pathways of COVID-19 transmission. Recent studies confirmed COVID-19 can be transmitted in three ways. Firstly, patients could have acquired the virus through contaminated touchable surfaces[5,6]. Hospitals, consequently, have been one of the major sources of the virus transmission. Other high-risk areas might also include building elevators, metro stations and work offices. Another way of transmitting COVID-19 is through the aerosolized viral particles[7]. Infected individuals could spread the virus by sneezing, talking, coughing and vomiting. In this regard, besides the areas that are mentioned above, schools, shopping malls, local markets and wholesale markets could also be the high-risk areas[8,9]. Latest studies also found fecal-oral could be another transmission pathway, due to the presence of COVID-19 in stool.[7,10-12] Therefore, high-rise buildings could be high-risk areas because large numbers of flats share the same sewer system. By far, there is no evidence that the airborne transmission of COVID-19 is possible. However, empirical studies found the airborne transmission of the severe acute respiratory syndrome (SARS)[13], other kind of coronavirus. Therefore, we should also consider areas that have degraded ventilation facility, such as aging buildings. In conclusion, eight kinds of BEs could be high-risk areas of infection in the city (see Data and Methods for details of all eight kinds of BEs). To understand how those BE play roles in transmitting COVID-19, the ideal way is to measure the interaction between the human beings and the contaminated BEs. However, due to the lack of such data, this study instead measures the confirmed cases in a walking distance to them. The rationale is if one lives geographically close to the contaminated BE, one should have higher chances interacting with the contamination or the infected individuals whom had interacted with it. Therefore, our hypothesis is *highly infection areas (or neighborhoods) should be geographically closer to the BEs that tend to spread COVID-19*. This study tries to provide empirical evidences to test the hypothesis. At the end of result section, we also provide two case studies that illustrate how residents of a neighborhood could receive the virus from surrounding BEs of infection risks.

This study collects 7,826 confirmed cases from 1,176 *xiaoqu* in Wuhan City, the epi-center of COVID-19. A *xiaoqu* is a group of buildings (or sometimes a single building) that share a common yard with some public space. Residents need to have a certified pass to enter their *xiaoqu*, so it resides a compact and dense community. We also collect the demographic information of each *xiaoqu*, along with its building ages, the green area ratio (GAR), floor area ratio (FAR), the closest hospitals, metro stations, parks, shopping centers, local food markets, wholesale markets and schools. We divide the *xiaoqu* in three groups according to the levels of infection (see Data and Methods for more details) and they are called high-infection *xiaoqu*, medium-infection *xiaoqu* and low-infection *xiaoqu*. To prove our hypothesis, we test whether the highly infected *xiaoqu* is significantly closer to the BE that are regarded as high-risk facilities than other kinds of *xiaoqu*. We present our findings in the result section.

# Data and method

### Data sources

The COVID-19 confirmed cases are derived from the daily outbreak bulletin released by the *xiaoqu* administrators (see in the Supplementary Materials (SM) for a sample of the bulletin). A total of 7,826 confirmed cases from 1,176 *xiaoqu* were collected, accounting for 15.7% of the total cases in Wuhan (till March 22)[14]. The spatial distribution is shown in Figure 1.

The demographic data of each *xiaoqu*, including the numbers of residents, the population density and the numbers of aged population (age >= 65 years old), is derived from *the Actual Population Data of 2014*[15]. Since the statistical unit is at community level which covers several *xiaoqu*, we calculate *xiaoqu's* population by the proportion of *xiaoqu's* residential land use area to its corresponding community's residential land use area. Similarly, we use the community's aged population rate to determine *xiaoqu's* aged population rate.

Eight BEs around the *xiaoqu*, including schools, parks, metro stations, hospitals, local food markets, wholesale markets, shopping malls and *xiaoqu* itself, are derived from point of interest (poi) in AutoNavi Map amap.com[16]. The BE in the *xiaoqu,* including floor area ratio (FAR), green area ratio (GRA), and building's age*,* is derived from the online real estate advertising agency Fang.com[17].

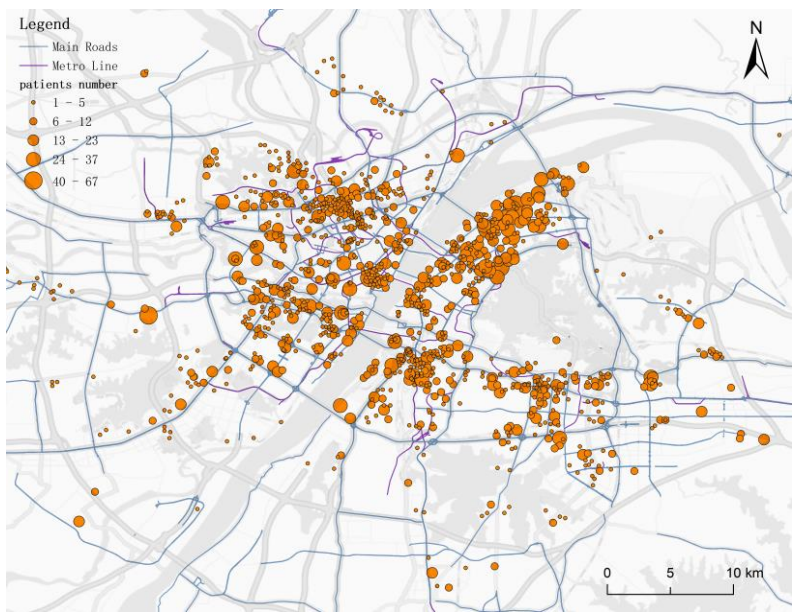

**Figure 1. the spatial distribution of sample confirmed cases in Wuhan.** The circle in orange indicates the infected *xiaoqu*, with the circle size proportional to the numbers of cases.

## Data characteristics

To ensure the collected confirmed cases are representative to the overall situation in Wuhan, we compare the temporal changes of the sample data against the official reported cases in Wuhan (see Figure 2). We found that temporal trend of the sample data aligning well with the reported cases, which proves the sample data is representative to the overall situation in Wuhan.

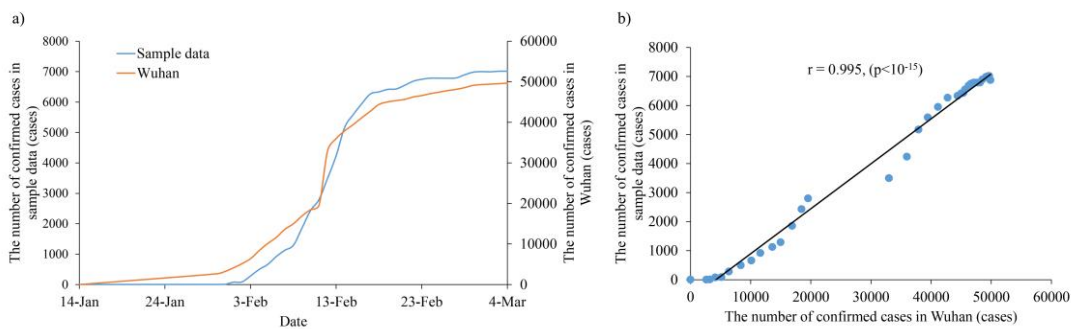

**Figure 2. temporal alignment between the sample data and the official reported cases in Wuhan.** a) the line chart shows the accumulated numbers of cases in the sample data (in blue) and the official reported cases of Wuhan (in orange) between January 14th and March 4th. The chart confirms the sample data reflects the temporal trends of the overall situation, where the outbreak started on

February 3rd, and the number of confirmed cases began to increase rapidly. On around February 19th, the outbreak has been under control. b) The points represent the numbers of sample cases and reported cases in a single day. The Pearson Coefficient between all reported cases and sample cases is r = 0.995 ($p<10^{-15}$).

**Estimates of the BE effects on the COVID-19 infection risks**

This study divides 1,176 *xiaoqu* into three groups based on the levels of infection (see Figure 3). Each group has approximately equal numbers of confirmed cases. We call them high-infection *xiaoqu*, medium-infection *xiaoqu* and low-infection *xiaoqu* (abbreviated as H-IX, M-IX and L-IX respectively). The Gini index is 50.3% (see Figure 3), which indicates the distribution of confirmed cases across the *xiaoqu* is extremely uneven. In details, the H-IX account for 8.6% of the overall samples, while the M-IX and L-IX account for 20.3% and 71.0% respectively. In terms of the distribution of confirmed cases, in H-IX, the maximum confirmed case is 67, the minimum is 17 and the average is 25.7. In M-IX the maximum confirmed case is 17, the minimum is 7 and the average is 10.9. In L-IX, the maximum confirmed case is 7, the minimum is 1 and the average is 3.1.

To estimate the BE effects, we use two different ways. Firstly, we measure whether there is a significant difference of distance between those three types of *xiaoqu* and their closest BE. For example, to estimate whether or not H-IX is closer to hospital compared to M- and L-IX, we compute the distance of each *xiaoqu* to its closest hospital and accordingly perform T-test of the distance distributions between all three groups. If the T-test fails to reject the null hypothesis (p > 0.05), we affirm no difference between the distance distributions. That means living close to a hospital one would not be more exposed to the infection risks of COVID-19. The other way is to determine whether there is significant difference between those three groups of *xiaoqu* in terms of the numbers of BE within a walking distance. In this method, we use buffer areas (r = [0.5, 1, … , 2km]) with each *xiaoqu* as the central point to compute the numbers of BE of the same kind, and accordingly compare the distributions between those three groups of *xiaoqu* using T-test. Similarly, if T-test fails to reject the null hypothesis, we found living closed to the corresponding BE would not expose the residents to the infection risk of COVID-19.

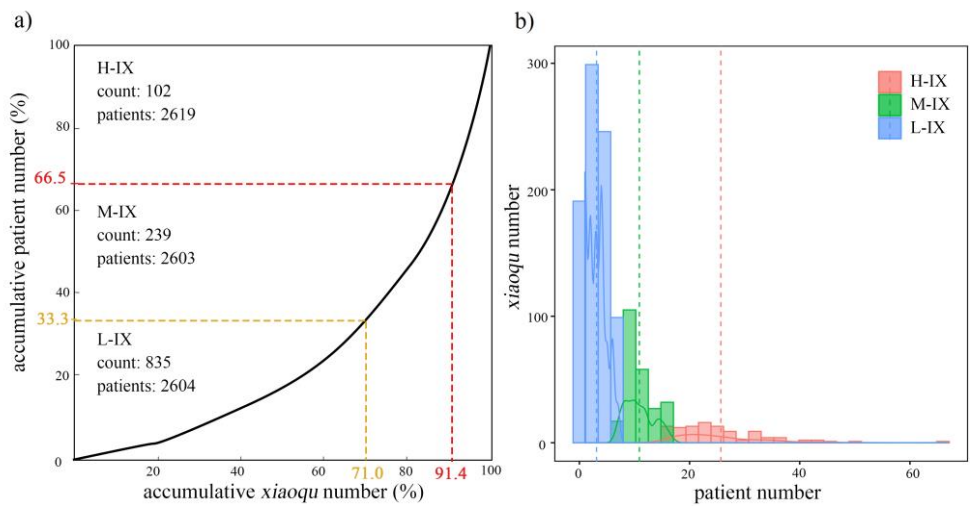

**Figure 3. the distribution of patients across three kinds of** *xiaoqu*. a) We use Lorenz curve to determine the thresholds of confirmed cases for grouping. Three groups H-, M- and L-IX, have approximately equal numbers of confirmed cases. b) the distributions of confirmed cases across three different groups.

# Result

## Demographics

Three demographic attributes are used to describe the sample *xiaoqu*: resident amount, population density and the percentage of aged population (age >= 65 years old). The statistics of all demographics and BE features across H-, M- and L-IX is displayed in table 1. According to Table 1, H-IX reside more residents on average. The average resident numbers of H-IX is 3,652, significantly larger than both H- (2,388) and L-IX (1,437). In terms of aged population ratio, H-IX reside an average aged population ratio at 14%, significantly larger than those of H- and L-IX (both are about 11%). There exists no significant difference between M- and L-IX in terms of aged population ratio (see Figure 4). Empirical studies found aged population are more vulnerable to COVID-19, so this study confirms that at the *xiaoqu* level[18,19].

**Table 1 statistics of BE features and the T-test result.**

| feature | Attribute | low-infection (L) | medium-infection (M) | high-infection (H) | all | T-test |
|---|---|---|---|---|---|---|
| demographics | resident number | 1437 | 2388 | 3653 | 1839 | L<M<H |
|  | density (person/ha) | 281 | 288 | 317 | 286 | L--M--H |
|  | aged pop ratio | 0.11 | 0.11 | 0.14 | 0.11 | M<H>L |
| external environment: distance to urban facilities (in meter) | school | 611 | 555 | 409 | 582 | M>H<L |
|  | park | 1605 | 1651 | 1214 | 1580 | M>H<L |
|  | metro station | 1313 | 1287 | 1020 | 1282 | M>H<L |
|  | hospital | 1363 | 1570 | 1158 | 1387 | M>H<L |
|  | food market | 486 | 403 | 284 | 452 | L>M>H |
|  | shopping mall | 1476 | 1415 | 1074 | 1429 | M>H<L |
|  | wholesale market | 3245 | 3488 | 2845 | 3259 | L--M>H |
| external environment: numbers of urban facilities inside a walking distance | school (1km) | 3.0 | 2.7 | 4.0 | 3.0 | M<H>L |
|  | park (1.5km) | 1.1 | 1.3 | 1.6 | 1.2 | M--L<H |
|  | metro station (1km) | 0.8 | 0.7 | 1.0 | 0.8 | M<H>L |
|  | hospital (1.5km) | 3.3 | 2.7 | 3.7 | 3.2 | L>M<H |
|  | food market | 1.5 | 1.6 | 2.2 | 1.6 | M<H>L |
|  | (0.5km) |  |  |  |  |  |
|  | mall (1.5km) | 2.4 | 1.9 | 2.6 | 2.3 | L<M--H |
|  | wholesale market (1.5km) | 1.6 | 1.0 | 1.1 | 1.4 | L<M--H |
| **internal environment** | green area ratio | 34.4 | 34.9 | 35.0 | 34.5 | L--M--H |
|  | floor area ratio | 2.6 | 2.6 | 2.5 | 2.6 | L--M--H |
|  | building year | 2007 | 2007 | 2006 | 2007 | L--M--H |

significate: P-value<0.05     > significantly larger; < significantly smaller; -- not significant

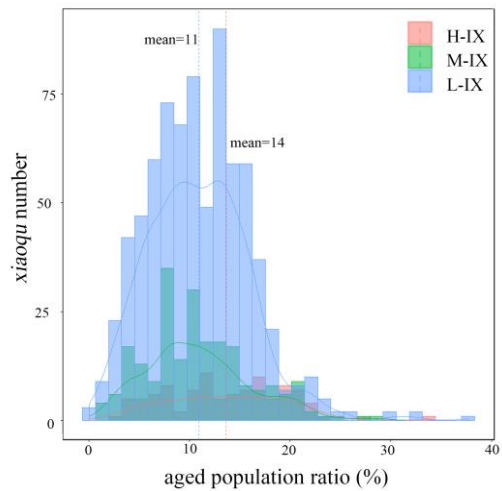

**Figure 4. Frequency distribution of aged population ratio across three groups.**

It is worth noting that population density does not show significant difference across three groups. Population density has always been used to imply the intensity of human interaction, through which COVID-19 transmits[20]. Therefore, one would hypothesize H-IX would be denser compared to M-IX and L-IX. This study does not support that hypothesis. One of the explanations would be the strict social distancing measures that were implemented from February 10th, 2020, reducing the interaction of residents inside the *xiaoqu*.

**Effects of external BE features.**

This study collects seven types of BEs that cover most of the urban activities like commute, entertainment, education, health care, and basic life services.

We found that H-IX are more closed to these urban facilities on average (see Table 1). The difference of mean values between H- and M- and L-IX are significant. For instance, the mean value of the shortest distance from school to H-IX is 409m, significantly smaller than that of H- (at 555m) and L-IX (at 611m). Especially for food markets, the average distances to H-IX is 284m, nearly half of the distances to L-IX (at 486m). Other BEs such as parks, metro stations, hospitals, shopping malls, wholesale markets show similar results, which are as depicted in table 1 and figure 5.

In terms of BE amount within a walking distance, we found that, H-IX can access more facilities in general (see Figure 6). Comparing between H-IX and L-IX, we found H-IX can access more schools, parks, metro stations and food markets than L-IX, no matter in which buffer. In terms of the accessibility to hospitals, H-IX can only access more hospitals than L-IX in the 0.5km walking distance. There is no significant difference in terms of the accessibility for wholesale markets and shopping malls between H-IX and L-IX.

Comparing H- and M-IX, H-IX can access more schools and metro stations, no matter in which distance category. In terms of parks, H-IX can access more in the distance of 2km and 2.5km. For food markets, between 0.5km and 2km, and for hospitals, between 0.5km and 1.5km, H-IX can significantly access more than M-IX. However, there is also no significant difference for wholesale markets and shopping malls.

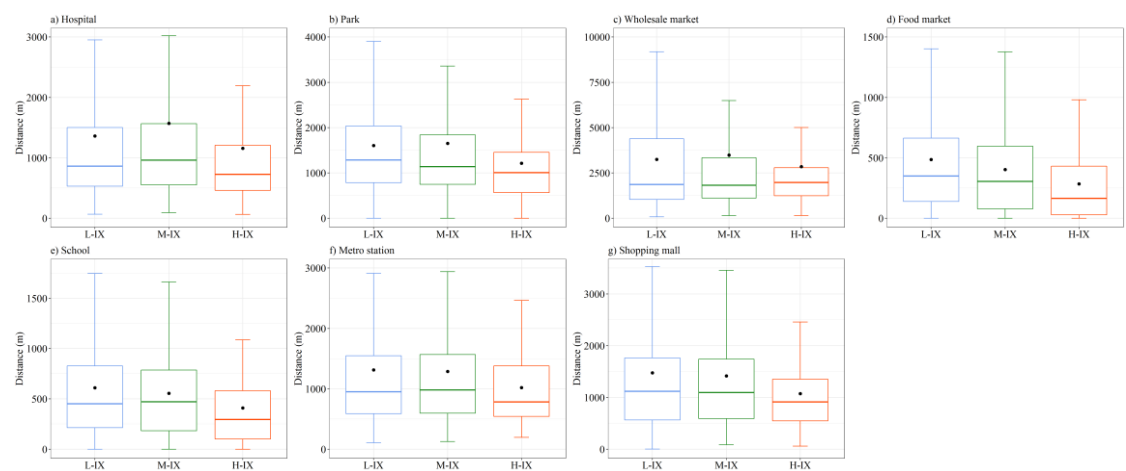

**Figure 5. Comparing shortest distances to seven BEs between H-, M- and L-IX.** H-IX are closer to seven types of BEs than M- and L-IX. These BEs include hospitals, parks, wholesale markets, food markets, schools, metro stations and shopping malls.

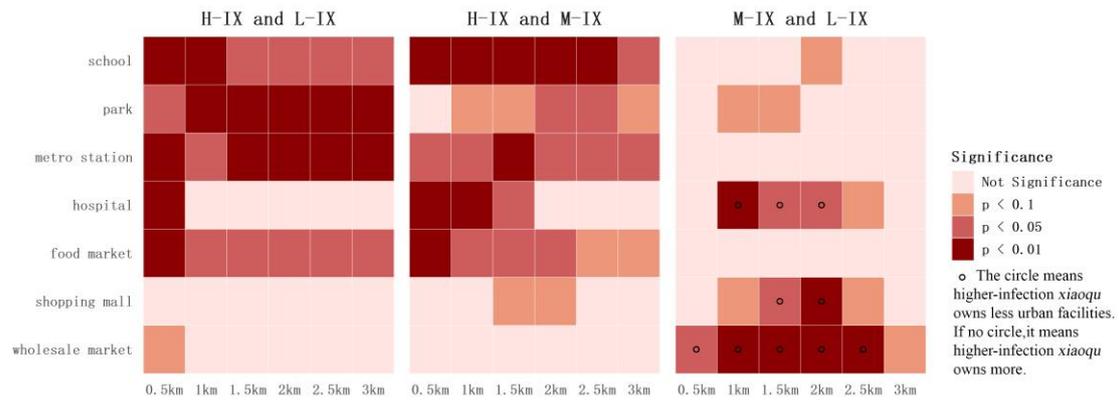

**Figure 6.** Comparing BE accessibility between H-, M- and L-IX within a walking distance.

### Effects of internal BE features.

Internal building environment is the feature of the *xiaoqu* itself, which includes the FAR, GAR and building year. FAR is the ratio of the whole architecture floor area to the land area, while GAR is the ratio of the green area to the whole land area. The former is an index for measuring how dense the community is and the latter is for measuring how many open spaces could be used for human outdoor activities or ecological protection. Building year is the year when the *xiaoqu* is constructed.

Table 1 addresses the average FAR and GAR of all *xiaoqu* are 2.6 and 34.5%, respectively. The average building year is 2007. There exists no significant difference among the three groups in all of these three features. Therefore, the internal BE take no effect on the spread of COVID-19.

### Two case studies of the H-IX

As can be seen from the findings discussed above, some BEs will increase the outbreak risks of the community. As the Huanan Seafood Wholesale Market has been regarded as the first place of the outbreak, its surrounding communities are the first to be threatened[21]. To address how these BE play roles in spreading virus to the surrounding communities, this study selects two examples of H-IX, the No.121 and Evergrande Oasis *xiaoqu*, trying to demonstrate the spread pathway.

**No.121 *xiaoqu***

There were 37 cases diagnosed in the No.121 *xiaoqu* by February 17, ranking the 8th in our samples. This *xiaoqu* is located in the central area of GangHua village street and belongs to the staff settlements of Wuhan Iron and Steel Corporation (WISCO), which was built in 2000 and resides a total of 3,714 families.

Two main roads that are full of restaurants and shops surround the *xiaoqu*. Besides, at the intersection of that two main roads are a wholesale meat market and a hospital. Wuhan Garden Science Park is located behind that hospital, about 200 meters away from the *xiaoqu*. In addition, there are four schools within 200 meters of the *xiaoqu*, causing the pedestrian congestion during the drop off and pick up hours of school. With these urban settings (see S2 of SM), almost all kinds of engaged social interactions, including shopping, schooling, medical services and recreation, intermingled in this crowded space, creating a perfect opportunity for virus transmission.

Additionally, the buildings in this *xiaoqu* are old, and most of the young and affluent residents have moved out of the *xiaoqu*. It has gradually become home of many senior people. As most of the residents used to work for WISCO, they have similar living habits and educational background, forming a very strong social bond between them. Once a person is infected, the virus can spread quickly.

**Evergrande oasis *xiaoqu***

There were 44 cases diagnosed in the Evergrande oasis *xiaoqu* by March 1, ranking the 5th in the sample data. This *xiaoqu* is located in Sino-French Eco-City, which is a suburban community. Built in 2011, evergrande oasis *xiaoqu* is a largely self-contained settlements, with kindergarten, primary school, clubs, supermarkets and shops. Residents' daily activities mainly take place inside the xiaoqu where all needed facilities are concentrated (see S3 of SM for the urban settings illustration).

This *xiaoqu* is a hub for public transportation, and the mass commuting flow increases the risk of infection. It is also close to the bus station and Xintian metro station. There are a number of buses to the CBD and the subway to the Wang Jiawan Business Circle takes only 10 minutes. In the crowded subways, it is easy for residents to infect the virus.

In addition, this *xiaoqu* is only one stop away from Tongji hospital. As a COVID-19 designated hospital, it has been treating a large number of infected patients every day since the outbreak of the epidemic and has undertaken the medical rescue mission in western Wuhan.

# Conclusions and discussions

This study relates the infection levels of *xiaoqu* to its demographics and surrounding BEs, to address the infection risks of different BEs (the tested BEs include *xiaoqu* itself). Firstly, in the aspect of demography, H-IXs own higher ratio of aged population than M- and L-IX on average. That means *xiaoqu* with more aged people are more likely to be infected by COVID-19. However, it shows no significant difference in population density across three groups. Secondly, from the perspective of external BEs, H-IXs are closer to seven types of urban facilities than M- and L-IX on average, including hospitals, parks, schools, food markets, wholesale markets, shopping malls and metro stations. In a walking distance, H-IX can access more of them except the wholesale markets and shopping malls. Thirdly, in the aspect of internal BEs, FAR, GAR and building year, it shows no significant difference across three groups.

These findings inform the policy-makers of two kinds of high-risk areas in the city, the aged *xiaoqu*s and the popular BEs that we tested in this study. Firstly, aged *xiaoqu*s are more susceptible to COVID-19 because aged people could be easily infected. More deliberate health care should be invested to such *xiaoqu*s, including daily body temperature monitoring and adding more physical excise space to improve aged group's immunity.

Additionally, social distancing measures should continually put in place the tested BEs in this study. For example, telecommuting and staggered peak travel should be encouraged to avoid the crowd of metro station. Temperature monitoring, even health self-report should be required when citizens access to these urban facilities. In the long-run, from the perspective of urban planning, public facilities should be located more evenly so that citizens could have equitable access while minimizing the infection risks.

# Endnote

1 Chinazzi M, Davis JT, Ajelli M, et al. The effect of travel restrictions on the spread of the 2019 novel coronavirus (COVID-19) outbreak. *Science*, 2020, eaba9757.

2 Prem K, Yang L, Russell TW, et al. The effect of control strategies to reduce social mixing on outcomes of the COVID-19 epidemic in Wuhan, China: a modelling study. *The Lancet Public Health*,2020,ISSN 2468-2667. DOI:10.1016/S2468-2667(20)30073-6.

3 WHO. Coronavirus disease (COVID-2019) situation reports. 2020. https://www.who.int/emergencies/diseases/novel-coronavirus-2019/situation-reports.

4 BBC. News. Pneumonia epidemic has no turning point - China's highest level decides to restore transportation security. 2020. https://www.bbc.com/zhongwen/simp/chinese-news51594007.

5 Rothan HA, Byrareddy SN. The epidemiology and pathogenesis of coronavirus disease (COVID-19) outbreak. *Journal of Autoimmunity*,2020, 102433, ISSN08968411. DOI:10.1016/j.jaut.2020.102433.

6 Sizun J , Yu MWN , Talbot P J. Survival of human coronaviruses 229E and OC43 in suspension and after drying on surfaces: a possible source of hospital-acquired infections. *journal of hospital infection*, 2000, 46(1):0-60.

7 Ong SWX , Tan YK , Chia PY , et al. Air, Surface Environmental, and Personal Protective Equipment Contamination by Severe Acute Respiratory Syndrome Coronavirus 2 (SARS-CoV-2) From a Symptomatic Patient. *The Journal of the American Medical Association*. DOI: 10.1001/jama.2020.3227.

8 Cai J, Sun W, Huang J, et al. Indirect virus transmission in cluster of COVID-19 cases, Wenzhou, China, 2020. *Emerg Infect Dis*. 2020 Jun [date cited]. DOI:10.3201/eid2606.200412.

9 Lu J, Gu J, Li K, et al. COVID-19 outbreak associated with air conditioning in restaurant, Guangzhou, China, 2020. *Emerg Infect Dis*. 2020 Jul [date cited]. DOI:10.3201/eid2607.200764.

10 Zhu N, Zhang D ,Wang W, et al. A Novel Coronavirus from Patients with Pneumonia in China, 2019. *N Engl J Med*. DOI: 10.1056/NEJMoa2001017.

11 Xiao F, Tang M, Zheng X,et al. Evidence for Gastrointestinal Infection of SARS-CoV-2. *Gastroenterology*, 2020, ISSN 0016-5085, DOI: 10.1053/j.gastro.2020.02.055.

12 Wu Y, Guo C, Tang L, et al. Prolonged presence of SARS-CoV-2 viral RNA in faecal samples. *Lancet Gastroenterol Hepatol 2020*.https://doi.org/10.1016/S2468-1253（20）30083-2.

13 Booth TF, Bill K , Nathalie B , et al. Detection of Airborne Severe Acute Respiratory Syndrome (SARS) Coronavirus and Environmental Contamination in SARS Outbreak Units. *Journal of Infectious Diseases*(9):9. DOI: 10.1086/429634.

14 Epidemic data in China: http://2019ncov.chinacdc.cn/2019-nCoV/.

15 Actual Population Database: The basic information database of the community residents established through the field survey. The database includes residents' housing, separation status, health status, actual employment status and so on.

16 Amap: https://www.amap.com/.

17 fang.com:https://wuhan.esf.fang.com/.

18 Verity R, Okell LC, Dorigatti I, et al. Estimates of the severity of coronavirus disease 2019: a model-based analysis. *Lancet Infect Dis 2020*. Published Online March 30, 2020.

19 The Novel Coronavirus Pneumonia Emergency Response Epidemiology Team. The epidemiological characteristics of an outbreak of 2019 novel coronavirus diseases (COVID-19) in China. *Chinese Journal of Epidemiology*, 2020, 41(2): 145-151.

20 Luís M A Bettencourt, José Lobo, Helbing D, et al. Growth, Innovation, Scaling, and the Pace of Life in Cities. *Proceedings of the National Academy of Sciences*, 2007, 104(17):7301-7306.

21 Huang C, Wang Y, Li X, et al. Clinical features of patients infected with 2019 novel coronavirus in Wuhan, China. *Lancet 2020*; 395: 497–506.

# Supplementary Materials

**S1. A sample of the bulletin released by the xiaoqu administrator.** The bulletin addresses the total numbers of confirmed cases, and sometimes the building number.

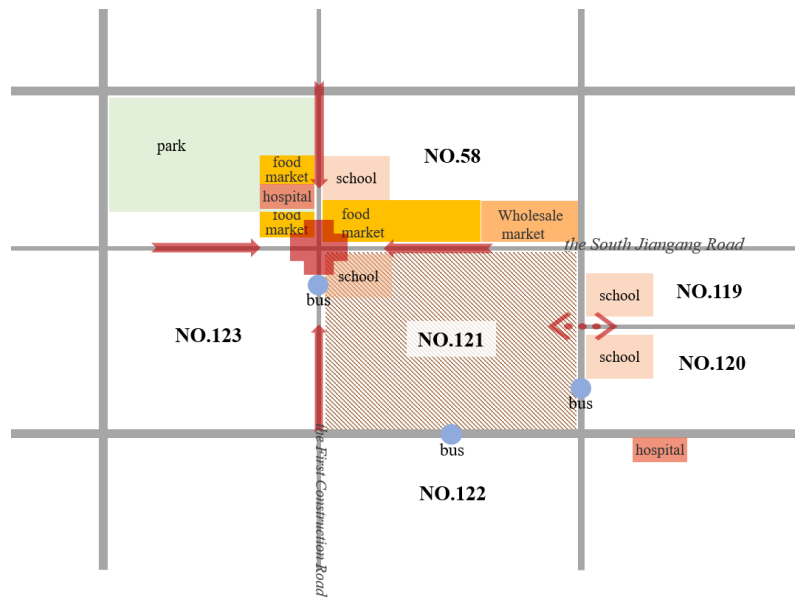

**S2. NO.121 *xiaoqu* and its surrounding urban Settings**.

**S3. Evergrande oasis *xiaoqu* and its surrounding urban Settings.**

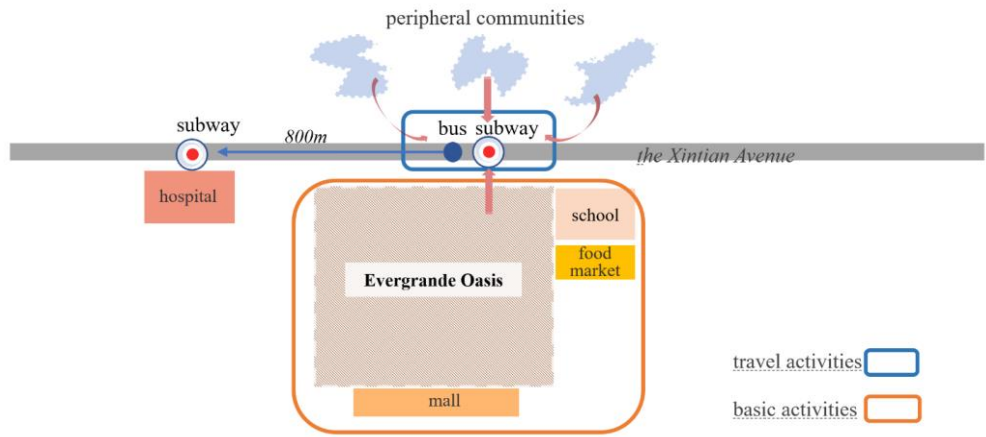